\documentclass[aps,prl,twocolumn,showpacs,groupedaddress]{revtex4}  % for review and submission 
\usepackage{graphicx}  % needed for figures
\usepackage{dcolumn}   % needed for some tables
\usepackage{bm}        % for math
\usepackage{amssymb}   % for math
\usepackage[dvips]{color}
\usepackage{ifthen}

\newboolean{color}
\setboolean{color}{false}      %%% use gray-level figures

\def \ud {{1 \over 2} }
\newcommand{\rpv}{\mbox{$\not \hspace{-0.10cm} R_p$}}
\newcommand{\MET}{\mbox{$\not \hspace{-0.10cm} E_T$}}
\newcommand{\ld}{$\lambda'_{211}$~}

\newcommand{\lqd}{\mbox{$LQ\bar{D}$}}

\begin{document}

% the following line is for submission 
\hspace{5.2in} \mbox{Fermilab-Pub-06-094-E}

\title{Search for resonant second generation slepton production at the Tevatron}
% LIST_OF_AUTHORS_R2.TEX                 4/27/06            
%
\author{                                                                      
%% names begin here                                                           
V.M.~Abazov,$^{36}$                                                           
B.~Abbott,$^{76}$                                                             
M.~Abolins,$^{66}$                                                            
B.S.~Acharya,$^{29}$                                                          
M.~Adams,$^{52}$                                                              
T.~Adams,$^{50}$                                                              
M.~Agelou,$^{18}$                                                             
J.-L.~Agram,$^{19}$                                                           
S.H.~Ahn,$^{31}$                                                              
M.~Ahsan,$^{60}$
% ...
\\                                                                            
\vskip 0.30cm                                                                 
\centerline{(D\O\ Collaboration)}                                             
\vskip 0.30cm                                                                 
}                                                                             
\affiliation{                                                                 
\centerline{$^{1}$Universidad de Buenos Aires, Buenos Aires, Argentina}       
\centerline{$^{2}$LAFEX, Centro Brasileiro de Pesquisas F{\'\i}sicas,         
                  Rio de Janeiro, Brazil}                                     
\centerline{$^{3}$Universidade do Estado do Rio de Janeiro,                   
                  Rio de Janeiro, Brazil}                                     
\centerline{$^{4}$Instituto de F\'{\i}sica Te\'orica, Universidade            
                  Estadual Paulista, S\~ao Paulo, Brazil}                     
\centerline{$^{5}$University of Alberta, Edmonton, Alberta, Canada,           
                  Simon Fraser University, Burnaby, British Columbia, Canada,}
\centerline{York University, Toronto, Ontario, Canada, and                    
                  McGill University, Montreal, Quebec, Canada}                
\centerline{$^{6}$Institute of High Energy Physics, Beijing,                  
                  People's Republic of China}                                 
\centerline{$^{7}$University of Science and Technology of China, Hefei,       
                  People's Republic of China}                                 
\centerline{$^{8}$Universidad de los Andes, Bogot\'{a}, Colombia}             
\centerline{$^{9}$Center for Particle Physics, Charles University,            
                  Prague, Czech Republic}                                     
\centerline{$^{10}$Czech Technical University, Prague, Czech Republic}        
\centerline{$^{11}$Center for Particle Physics, Institute of Physics,         
                   Academy of Sciences of the Czech Republic,                 
                   Prague, Czech Republic}                                    
\centerline{$^{12}$Universidad San Francisco de Quito, Quito, Ecuador}        
\centerline{$^{13}$Laboratoire de Physique Corpusculaire, IN2P3-CNRS,         
                   Universit\'e Blaise Pascal, Clermont-Ferrand, France}      
\centerline{$^{14}$Laboratoire de Physique Subatomique et de Cosmologie,      
                   IN2P3-CNRS, Universite de Grenoble 1, Grenoble, France}    
\centerline{$^{15}$CPPM, IN2P3-CNRS, Universit\'e de la M\'editerran\'ee,     
                   Marseille, France}                                         
\centerline{$^{16}$IN2P3-CNRS, Laboratoire de l'Acc\'el\'erateur              
                   Lin\'eaire, Orsay, France}                                 
\centerline{$^{17}$LPNHE, IN2P3-CNRS, Universit\'es Paris VI and VII,         
                   Paris, France}                                             
\centerline{$^{18}$DAPNIA/Service de Physique des Particules, CEA, Saclay,    
                   France}                                                    
\centerline{$^{19}$IReS, IN2P3-CNRS, Universit\'e Louis Pasteur, Strasbourg,  
                    France, and Universit\'e de Haute Alsace,                 
                    Mulhouse, France}                                         
\centerline{$^{20}$Institut de Physique Nucl\'eaire de Lyon, IN2P3-CNRS,      
                   Universit\'e Claude Bernard, Villeurbanne, France}         
\centerline{$^{21}$III. Physikalisches Institut A, RWTH Aachen,               
                   Aachen, Germany}                                           
\centerline{$^{22}$Physikalisches Institut, Universit{\"a}t Bonn,             
                   Bonn, Germany}                                             
\centerline{$^{23}$Physikalisches Institut, Universit{\"a}t Freiburg,         
                   Freiburg, Germany}                                         
\centerline{$^{24}$Institut f{\"u}r Physik, Universit{\"a}t Mainz,            
                   Mainz, Germany}                                            
\centerline{$^{25}$Ludwig-Maximilians-Universit{\"a}t M{\"u}nchen,            
                   M{\"u}nchen, Germany}                                      
\centerline{$^{26}$Fachbereich Physik, University of Wuppertal,               
                   Wuppertal, Germany}                                        
\centerline{$^{27}$Panjab University, Chandigarh, India}                      
\centerline{$^{28}$Delhi University, Delhi, India}                            
\centerline{$^{29}$Tata Institute of Fundamental Research, Mumbai, India}     
\centerline{$^{30}$University College Dublin, Dublin, Ireland}                
\centerline{$^{31}$Korea Detector Laboratory, Korea University,               
                   Seoul, Korea}                                              
\centerline{$^{32}$SungKyunKwan University, Suwon, Korea}                     
\centerline{$^{33}$CINVESTAV, Mexico City, Mexico}                            
\centerline{$^{34}$FOM-Institute NIKHEF and University of                     
                   Amsterdam/NIKHEF, Amsterdam, The Netherlands}              
\centerline{$^{35}$Radboud University Nijmegen/NIKHEF, Nijmegen, The          
                  Netherlands}                                                
\centerline{$^{36}$Joint Institute for Nuclear Research, Dubna, Russia}       
\centerline{$^{37}$Institute for Theoretical and Experimental Physics,        
                   Moscow, Russia}                                            
\centerline{$^{38}$Moscow State University, Moscow, Russia}                   
\centerline{$^{39}$Institute for High Energy Physics, Protvino, Russia}       
\centerline{$^{40}$Petersburg Nuclear Physics Institute,                      
                   St. Petersburg, Russia}                                    
\centerline{$^{41}$Lund University, Lund, Sweden, Royal Institute of          
                   Technology and Stockholm University, Stockholm,            
                   Sweden, and}                                               
\centerline{Uppsala University, Uppsala, Sweden}                              
\centerline{$^{42}$Physik Institut der Universit{\"a}t Z{\"u}rich,            
                   Z{\"u}rich, Switzerland}                                   
\centerline{$^{43}$Lancaster University, Lancaster, United Kingdom}           
\centerline{$^{44}$Imperial College, London, United Kingdom}                  
\centerline{$^{45}$University of Manchester, Manchester, United Kingdom}      
\centerline{$^{46}$University of Arizona, Tucson, Arizona 85721, USA}         
\centerline{$^{47}$Lawrence Berkeley National Laboratory and University of    
                   California, Berkeley, California 94720, USA}               
\centerline{$^{48}$California State University, Fresno, California 93740, USA}
\centerline{$^{49}$University of California, Riverside, California 92521, USA}
\centerline{$^{50}$Florida State University, Tallahassee, Florida 32306, USA} 
\centerline{$^{51}$Fermi National Accelerator Laboratory,                     
            Batavia, Illinois 60510, USA}                                     
\centerline{$^{52}$University of Illinois at Chicago,                         
            Chicago, Illinois 60607, USA}                                     
\centerline{$^{53}$Northern Illinois University, DeKalb, Illinois 60115, USA} 
\centerline{$^{54}$Northwestern University, Evanston, Illinois 60208, USA}    
\centerline{$^{55}$Indiana University, Bloomington, Indiana 47405, USA}       
\centerline{$^{56}$University of Notre Dame, Notre Dame, Indiana 46556, USA}  
\centerline{$^{57}$Purdue University Calumet, Hammond, Indiana 46323, USA}    
\centerline{$^{58}$Iowa State University, Ames, Iowa 50011, USA}              
\centerline{$^{59}$University of Kansas, Lawrence, Kansas 66045, USA}         
\centerline{$^{60}$Kansas State University, Manhattan, Kansas 66506, USA}     
\centerline{$^{61}$Louisiana Tech University, Ruston, Louisiana 71272, USA}   
\centerline{$^{62}$University of Maryland, College Park, Maryland 20742, USA} 
\centerline{$^{63}$Boston University, Boston, Massachusetts 02215, USA}       
\centerline{$^{64}$Northeastern University, Boston, Massachusetts 02115, USA} 
\centerline{$^{65}$University of Michigan, Ann Arbor, Michigan 48109, USA}    
\centerline{$^{66}$Michigan State University,                                 
            East Lansing, Michigan 48824, USA}                                
\centerline{$^{67}$University of Mississippi,                                 
            University, Mississippi 38677, USA}                               
\centerline{$^{68}$University of Nebraska, Lincoln, Nebraska 68588, USA}      
\centerline{$^{69}$Princeton University, Princeton, New Jersey 08544, USA}    
\centerline{$^{70}$State University of New York, Buffalo, New York 14260, USA}
\centerline{$^{71}$Columbia University, New York, New York 10027, USA}        
\centerline{$^{72}$University of Rochester, Rochester, New York 14627, USA}   
\centerline{$^{73}$State University of New York,                              
            Stony Brook, New York 11794, USA}                                 
\centerline{$^{74}$Brookhaven National Laboratory, Upton, New York 11973, USA}
\centerline{$^{75}$Langston University, Langston, Oklahoma 73050, USA}        
\centerline{$^{76}$University of Oklahoma, Norman, Oklahoma 73019, USA}       
\centerline{$^{77}$Oklahoma State University, Stillwater, Oklahoma 74078, USA}
\centerline{$^{78}$Brown University, Providence, Rhode Island 02912, USA}     
\centerline{$^{79}$University of Texas, Arlington, Texas 76019, USA}          
\centerline{$^{80}$Southern Methodist University, Dallas, Texas 75275, USA}   
\centerline{$^{81}$Rice University, Houston, Texas 77005, USA}                
\centerline{$^{82}$University of Virginia, Charlottesville,                   
            Virginia 22901, USA}                                              
\centerline{$^{83}$University of Washington, Seattle, Washington 98195, USA}  
}                                                                             
%end                                                                          
  % input Dzero author list
\date{May 3, 2006}

%% Abstract -------------------------------------------------------------------
\begin{abstract}
  We present a search for supersymmetry in the $R$-parity violating resonant
  production and decay of smuons and muon-sneutrinos in the channels
  $\tilde{\mu}\to\tilde{\chi}^0_1\,\mu$,
  $\tilde{\mu}\to\tilde{\chi}^0_{2,3,4}\,\mu$, and
  $\tilde{\nu}_\mu\to\tilde{\chi}^\pm_{1,2}\,\mu$.   We analyzed
  $0.38$~fb$^{-1}$ of integrated luminosity collected between April 2002 and
  August 2004 with the D0 detector at the Fermilab Tevatron Collider. The
  observed number of events is in agreement with the standard model
  expectation,  and we calculate 95\%~C.L. limits on the slepton
  production cross section times branching fraction to gaugino plus muon, as a
  function of slepton and gaugino masses.  In the framework of minimal
  supergravity, we set limits on the coupling parameter $\lambda'_{211}$,
  extending significantly previous results obtained in Run I of the Tevatron
  and at the CERN LEP collider.
\end{abstract}
\maketitle 

%% Theory ---------------------------------------------------------------------

% $R$-parity
Supersymmetry (SUSY) predicts the existence of a new particle for every
standard model (SM) particle, differing by half a unit in spin.  The quantum
number $R$-parity~\cite{farrar}, defined as $R = (-1)^{3B+L+2S}$, where $B$,
$L$ and $S$ are the baryon, lepton and spin quantum numbers, is $+1$ for SM
and $-1$ for SUSY particles. Often $R$-parity is assumed to be conserved,
which leaves the lightest supersymmetric particle (LSP) stable. However, SUSY 
does not require $R$-parity conservation.

% $R$-parity violating Lagrangian
If $R$-parity violation (\rpv) is allowed, the following trilinear and bilinear
terms appear in the superpotential~\cite{rpv}:
\begin{eqnarray}
W_{\scriptsize \rpv}&=& \ud \lambda _{ijk} L_i^{\alpha} L_j^{\beta} \bar{E}_k
          +\lambda ' _{ijk} L_i^{\alpha} Q_j^{\beta} \bar{D}_k \nonumber\\
        &&  + \ud \lambda '' _{ijk} \bar{U}_i^{\xi} \bar{D}_j^{\psi} \bar{D}_k^{\zeta}
	  + \mu_i L_i H_1 
\label{super}
\end{eqnarray}
where $L$ and $Q$ are the lepton and quark SU(2) doublet superfields and $E$, $U$,
$D$ denote the singlet fields. The indices have the following meaning:  $i,j,k =
1,2,3$ = family index; $\alpha, \beta = 1,2 $ = weak isospin  index; $\xi,
\psi, \zeta = 1,2,3 $ = color index. 
The coupling strengths are given by the Yukawa coupling constants
$\lambda, \lambda'$ and  $\lambda''$. The last term, $\mu_i L_i H_1$, mixes
the lepton and the Higgs superfields. 
The $\lambda$ and $\lambda'$ couplings give rise to final states with multiple
leptons, which provide excellent signatures at the Tevatron.
   
% L violation and indirect limit from lepton universality
In the following we assume that only $\lambda'_{211}$ can be non-zero. This
implies (muon) lepton number violation. The \rpv\ coupling constants are  already
constrained by low-energy experiments, in particular $\lambda'_{211} < 0.059 \cdot
m_{\tilde q}/100 \, \mathrm{GeV}$~\cite{referbound}. 
For the squark masses $m_{\tilde q}$ kinematically accessible at the Tevatron, this limit on
$\lambda'_{211}$ is significantly improved by the present analysis.

% Existing Limits from previous searches
The D0 Collaboration searched for resonant slepton production in
Run~I~\cite{d0run1}. The H1 experiment at DESY searched for resonant squark
production~\cite{h1squark} in the framework  of $R$-parity violating
supersymmetry and  published limits on the couplings $\lambda'_{1jk}$. The
combined limits from the LEP collider at CERN \cite{barbier}, assuming
$R$-parity violating decay via \lqd-couplings, are $m$($\tilde{\chi}^0_1$)$\ge
39$~GeV, $m$($\tilde{\chi}^\pm_1$)$\ge 103$~GeV, $m$($\tilde{\nu}_\mu$)$\ge 78$~GeV
and $m$($\tilde{\mu}$)$\ge 90$~GeV. 

% Production
At $p\bar{p}$ colliders, an initial $q\bar{q}$ pair can produce a single
smuon or muon-sneutrino assuming a non-zero $\lambda'_{211}$ coupling.
The $s$-channel production is dominant and depends on the value of this 
coupling. The contributions of the $t$ and $u$ channels are negligible compared to the
resonant $s$ channel~\cite{deliot}.

% Decay
The slepton can then decay into a lepton and a gaugino without violating
$R$-parity. The $\lambda'_{211}$ coupling allows neutralino decays via virtual
sparticles (such as muon-sneutrinos, smuons and squarks) into two $1^{\rm st}$
generation quarks and one $2^{\rm nd}$ generation lepton.  The
$\tilde{\chi}^0_1$ decay branching fractions as predicted by mSUGRA, with the
ratio of the Higgs expectation values $\tan\beta=5$, the sign of the Higgsino
mass parameter $\mu<0$, and the common trilinear scalar coupling $A_0=0$, are
assumed, leading to BR($\tilde{\chi}^0_1\to\mu q_1\bar{q}_1'$) $\approx$
BR($\tilde{\chi}^0_1\to\nu_\mu q_1\bar{q_1}$).  In this analysis, the value of
\ld is always larger than $0.03$, therefore the corresponding decay length is
negligible.  The dominant slepton intermediate decays as well as the
corresponding final states are indicated in Table~\ref{table:sl_decays}. 

\begin{table}[h]
\begin{ruledtabular}
\begin{tabular}{llll}
 &$\tilde{l}$ decay channel & Dominant final states \\ \hline
 &$\tilde \mu \;      \to \;  \tilde \chi^{0}   \, \mu	   $ & 2~$\mu$, 2~jets \\
 &$\tilde \mu \;      \to \;  \tilde \chi^{\pm} \, \nu_\mu $ & 1~$\mu$, \MET, 4~jets\\
 &$\tilde{\nu}_\mu \; \to \;  \tilde \chi^{0}   \, \nu_\mu $ & 1~$\mu$, \MET, 2~jets\\
 &$\tilde{\nu}_\mu \; \to \;  \tilde \chi^{\pm} \, \mu	   $ & 2~$\mu$, 4~jets \\
\end{tabular}
\end{ruledtabular} 
\caption{Smuon and muon-sneutrino decay channels: the final states correspond
	 to $\tilde{\chi}^0_1\to\mu q_1\bar{q}_1'$ and $\tilde{\chi}^\pm_1\to
	 qq'\tilde{\chi}^0_1$.}
\label{table:sl_decays}
\end{table}
Because of the challenging multi-jet QCD environment and the advantage of the ability to
reconstruct the neutralino and smuon masses, at least two muons were required in the final
state. This leaves the three channels (i)~$\tilde{\mu}\to\tilde{\chi}^0_1\,\mu$,
(ii)~$\tilde{\mu}\to\tilde{\chi}^0_{2,3,4}\,\mu$, and
(iii)~$\tilde{\nu}_\mu\to\tilde{\chi}^\pm_{1,2}\,\mu$ which are analyzed independently. 
The analysis is insensitive to events where the $\tilde{\chi}^0_1$  decays into
$\nu_{\mu}\bar{q}q'$ and where no second muon is created in the cascade.

The data for this analysis were recorded by the D0 detector between April 2002 and
August 2004 at a center-of-mass energy of 1.96 TeV. The integrated luminosity
corresponds to $380\pm 25$~pb$^{-1}$. 

%%Detector section------------------------------------------------------------- 

\begin{figure*}[!t]
\begin{tabular}{ccc}
  \includegraphics[scale=1.0]{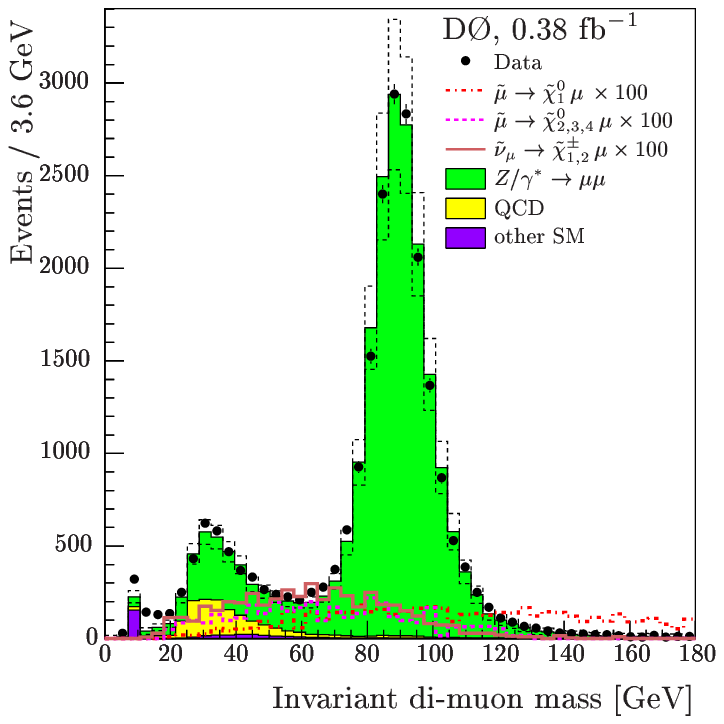} & \,\,\,\,\,\,&
  \includegraphics[scale=1.0]{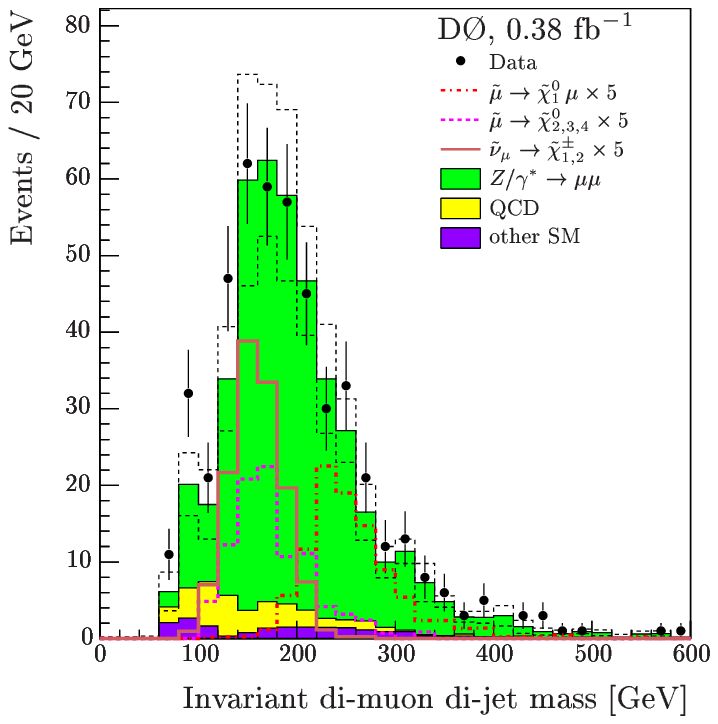} 
\end{tabular}
\caption{\label{fig:prslmet} Invariant di-muon mass in the two-muon sample
	 (a) and reconstructed 4-body mass of two muons and two jets
	 (b). The cascade decays in channels (ii) and (iii) lead to less
	 energy per particle, thus lower invariant masses. The signal
	 expectation for the point with $m_{\tilde{l}}=260$~GeV and
	 $m_{\tilde{\chi}^0_1}=100$~GeV is scaled in plot (a) by a factor of
	 $100$ and in plot (b) by a factor of $5$. The dominant SM
	 background is $Z/\gamma^*\to\mu\mu$; other SM backgrounds are
	 $Z/\gamma^*\to\tau\tau$;
	 $WW$, $WZ$, $ZZ$, $t\bar{t}$ and $\Upsilon$ production. The total SM
	 uncertainty is shown as a dashed black line. The data is
	 in good agreement with the SM expectation.}
\end{figure*}
The D0 detector~\cite{d0det} has a central tracking system consisting of a 
silicon microstrip tracker and a central fiber tracker,  both located within a
2~T superconducting solenoidal  magnet, with designs optimized for tracking
and  vertexing at pseudorapidities $|\eta|<3$ and $|\eta|<2.5$, respectively. A
liquid-argon and uranium calorimeter has a  central section covering
pseudorapidities $|\eta| \lesssim 1.1$, and two end calorimeters that extend
coverage  to $|\eta|\approx 4.2$.  The muon system covering $|\eta|<2.0$ 
consists of a layer of tracking detectors and scintillation trigger counters in
front of 1.8~T iron toroids, followed by two similar layers behind the toroids. 
The first level of the trigger (level 1) is based on fast information from
the tracking, calorimetry, and muon systems. At the next trigger stage (level
2), the rate is reduced further. These first two levels of
triggering rely mainly on hardware and firmware. The final level of the
trigger, level 3, with access to the full event information, uses software
algorithms to reduce the rate to tape to $50$~Hz.

The signal was simulated with {\sc susygen}~\cite{susygen}. The leading-order
{\sc susygen} signal cross sections have been multiplied by higher order,
slepton-mass dependent QCD-correction factors~\cite{inder} of size
$1.4-1.5$ calculated with the CTEQ6M~\cite{cteq} parton distribution functions
(PDFs). The influence of the PDF uncertainty on the cross section is $3\%-6\%$,
estimated from the CTEQ6M error functions.  The influence of the
renormalization scale and the factorization scale $\mu_F$ is less than $5\%$
for all slepton masses below 500~GeV, if $m(\tilde{l})/2\leq\mu_F\leq2\cdot
m(\tilde{l})$~\cite{maike}. 

The dominant background is inclusive production of $Z/\gamma^*\to\mu\mu$.  It
was simulated with the {\sc pythia}~\cite{pythia} Monte Carlo (MC) generator
and normalized using the predicted next-to-next-to leading order cross section~\cite{dy},
calculated with the CTEQ6 PDFs.  All other SM processes contribute only
slightly to the total background as seen in Fig.~\ref{fig:prslmet}. These
contributions were simulated using the {\sc pythia} and {\sc alpgen} generators
and normalized using next-to leading order cross section predictions calculated
using CTEQ6M PDFs.  All MC events were passed through a detailed detector
simulation based on {\sc geant} \cite{geant}, followed by the reconstruction
program used for data.

Events were collected with di-muon triggers requiring at least two muons at
level~1. At level~3 at least
one track or one muon with a varying transverse momentum $p_T$ threshold of
typically $5-15$~GeV was required. To account for the trigger effects,
simulated events were weighted using efficiencies determined from the data.

All events were required to contain two muons. One of the muons was required
to have $p_T>15$~GeV, and the second muon was required to have
$p_T>8$~GeV.  A central track match was required for both muons. 
The muons in the signal are expected to be isolated. We define muons as
``loose'' (``tight'') isolated, if the sum of the $p_T$ of the tracks in a
cone with radius $R_{\text{cone}}=\sqrt{\Delta\phi^2+\Delta\eta^2}=0.5$,
where  $\eta=-\ln \tan \frac{\theta}{2}$ is the pseudorapidity and $\theta$
is the azimuthal angle, around the muon direction is less than $10$~GeV
($2.5$~GeV), and the sum of the transverse energies of the calorimeter cells
in a hollow cone ($0.1 \le R_{\rm cone} \le 0.4)$) is less than $10$~GeV ($2.5$~GeV).  Both selected
muons were required to pass the tight isolation
requirement. The invariant di-muon mass
distribution of this di-muon sample is shown in Fig.~\ref{fig:prslmet}a.
At least two jets with transverse momentum $p_T \ge 15$~GeV and
reconstructed with a cone algorithm
($R_{\text{cone}}=0.5$)~\cite{Blazey:2000qt} were required. Only jets within $|\eta|<2.0$
were used. The reconstructed slepton mass with two muons and two jets is
shown in Fig.~\ref{fig:prslmet}b. The event selection is summarized in
Table~\ref{table:cutflow}.
\begin{table}[!b]
\begin{ruledtabular}
\begin{tabular}{lrr@{$\,\pm$}l@{$\,\pm$}rrr}
  Cut			   & Data  & \multicolumn{3}{c}{SM expectation}	   & 
  \multicolumn{2}{c}{Signal eff.}\\\hline%\rule{0mm}{2.5ex}%
  2$\mu$ selection 	   & 23206 &  22700  &70  &2900 & \multicolumn{2}{c}{5.5$\%$  $\pm$ 0.7$\%$} \\
  $p_T$ jet$_1 >$~15 GeV   &  3852 &   3760  &40  &560  & \multicolumn{2}{c}{4.8$\%$  $\pm$ 0.6$\%$} \\
  $p_T$ jet$_2 >$~15 GeV   &   475 &    430  &10  & 80  & \multicolumn{2}{c}{2.4$\%$  $\pm$ 0.3$\%$} \\
\end{tabular}
\end{ruledtabular} 
\caption{ Expected and observed events at different stages of the event selection. The signal
	  efficiency is given for the point with $m_{\tilde{l}}=260$~GeV and
	  $m_{\tilde{\chi}^0_1}=100$~GeV with respect to the total slepton
	  production. The first uncertainty on the SM expectation is statistical, the second
	  is due to systematics.}
\label{table:cutflow}
\end{table}
\begin{figure*}[!t]
   \begin{tabular}{lcr}
      \ifthenelse{\boolean{color}}
        {\includegraphics[scale=1.0]{figure_2a_color.eps}}
        {\includegraphics[scale=1.0]{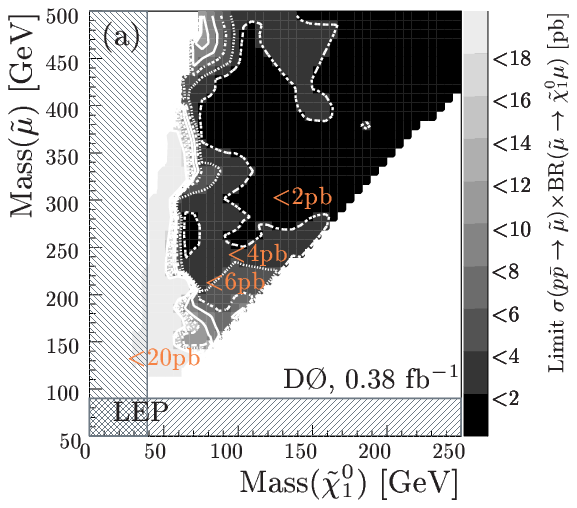}}
      \,\,\,&
      \ifthenelse{\boolean{color}}
        {\includegraphics[scale=1.0]{figure_2b_color.eps}}
        {\includegraphics[scale=1.0]{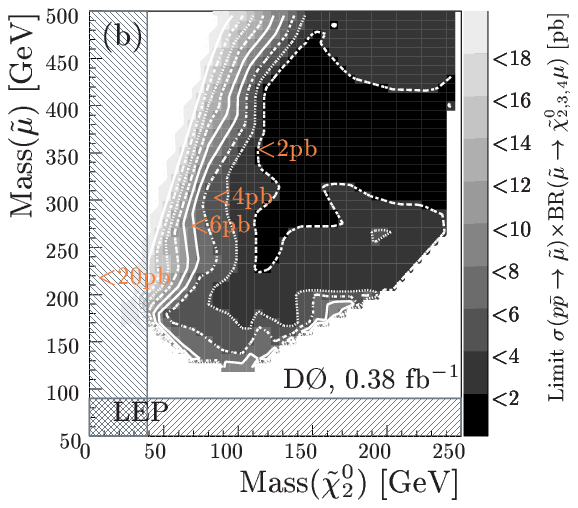}}
       \,\,\,&
      \ifthenelse{\boolean{color}}
        {\includegraphics[scale=1.0]{figure_2c_color.eps}}
        {\includegraphics[scale=1.0]{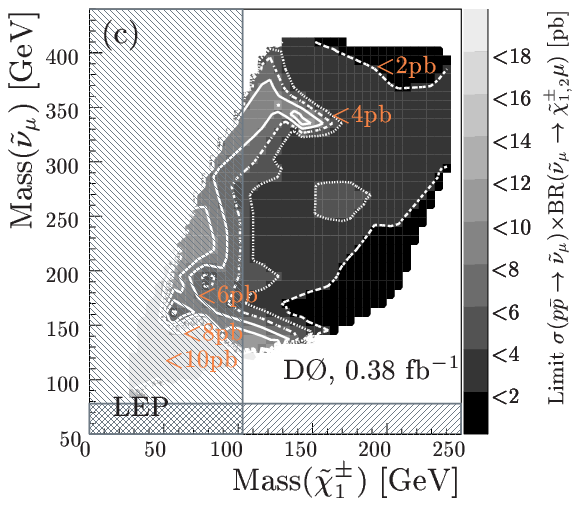}}
       \end{tabular}
  \caption{\label{fig:limxsecch1} $95\%$ C.L. limit on slepton production cross
	   section times branching fraction to gaugino plus muon for the
	   channels (i)~$\tilde{\mu}\to\tilde{\chi}^0_1\,\mu$ (a),
	   (ii)~$\tilde{\mu}\to\tilde{\chi}^0_{2,3,4}\,\mu$ (b) and 
	   (iii)~$\tilde{\nu}_\mu\to\tilde{\chi}^\pm_{1,2}\,\mu$ (c) as a
	   function of slepton and gaugino masses. The darkest region
	   corresponds to a cross section of less that $2$~pb. Successively
	   lighter regions have successively higher limits.
	   }
\end{figure*}
\begin{figure}[h]
     \begin{tabular}{c}
      \ifthenelse{\boolean{color}}
        {\includegraphics[scale=1.0]{figure_3_color.eps}}
        {\includegraphics[scale=1.0]{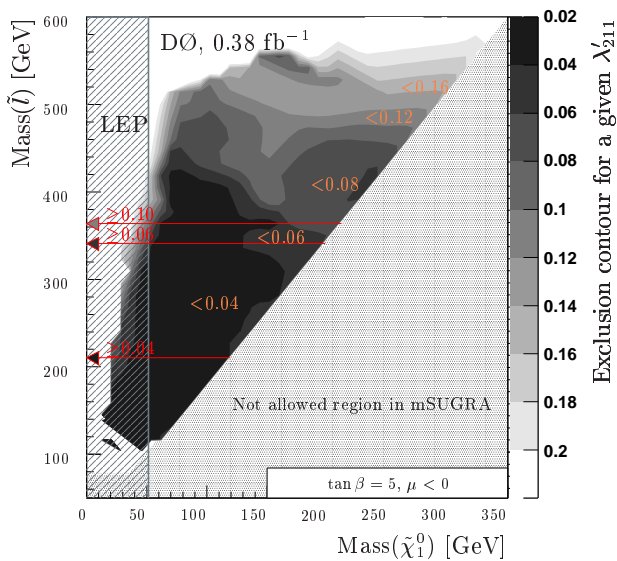}}
     \end{tabular}
  \caption{\label{fig:limldcomb} $95\%$ C.L. exclusion contour on $\lambda'_{211}$ couplings
           within the mSUGRA framework for $\tan\beta=5$ and $\mu<0$. The arrows indicate
	   %%absolute 
	   limits on the slepton mass $\tilde{l}$, for a given coupling $\lambda'_{211}$. }
\end{figure}

Background from multi-jet QCD events was extracted from data using loose muon isolation requirements. This QCD
enriched data sample was scaled to match the data in a signal free region. At least one isolation criterion with
respect to other energy depositions in the calorimeter or to other tracks must not be tight for at least one muon to
separate QCD and the data sample. 

%% 2D cut final selection ---------------------------------------------------------------------------
Two-dimensional selection requirements in planes spanned by the reconstructed
$\tilde{l}$ and $\tilde{\chi}$ candidate masses, the invariant di-muon and
di-jet masses, and the sums of muon momenta and jet momenta were used to
separate the signal $s$ from SM backgrounds $b$. The selection requirements
were chosen so that the {\it signal efficiency $\times$ signal purity}
$\propto \frac{s}{\sqrt{s+b}}$ of a specific cut, applied on a training
sample, was maximized. The selection requirements were optimized for each
(slepton mass, gaugino mass) combination (117 in total).

In the $\tilde{\mu}\to\tilde{\chi}^0_1\,\mu$ analysis (i), the slepton mass was
reconstructed with the two leading muons and the two jets. 
In the signal MC, the leading muon usually originates from the slepton
decay vertex. The neutralino mass was therefore reconstructed with both jets
and the next-to-leading muon. 

Hadronic decays of vector bosons from the gaugino cascade to $\tilde{\chi}^0_1$
can lead to additional jets in channels (ii) and (iii).  A simple likelihood
was calculated for each combination to reconstruct a vector boson and the
neutralino candidate mass. 
The slepton mass was reconstructed from all jets with $E_T>15$~GeV and
the two leading muons.

After the optimization, for the point with $m_{\tilde{l}}=260$~GeV and
$m_{\tilde{\chi}^0_1}=100$~GeV, we find $14/28/8$ events in the data while
$11.9\pm 2.1^{+1.5}_{-1.6}\ /\ 25.4\pm 3.2^{+6.7}_{-4.2}\ /\ 6.5\pm
1.6^{+2.0}_{-1.2}$ events are expected from SM backgrounds for the three
channels, respectively, with a typical signal efficiency of up to $2\%$. For all
$117$ mass combinations, the data is in agreement with the SM expectation
throughout the entire event selection range.

%% systematic errors ---------------------------------------------------------------------
The systematic uncertainties from different sources were added in quadrature.
For the limit calculation, the total systematic uncertainties of the
background and signal samples were taken to be $100\%$ correlated. A summary
of the uncertainties is given in Table \ref{table:syserrors} with their
contributions to the two muon and two jet sample.
\begin{table}[h]
\begin{ruledtabular}
\begin{tabular}{lcc}
Uncertainty		       & Background    & Signal \\ \hline 
Jet energy scale	       &  13.7\%     & 2 -- 26\%  \\
Muon ID                        &   7.8\%     & 8 --14\%   \\
Luminosity (does not apply to QCD) &   5.5\%  & 6.5\%  \\
Trigger efficiency  	       &   5.2\%     & 4 -- 9\%   \\
MC $\sigma$, $K$-factor, PDF   &   3.7\%     &  5\%	     \\
QCD background estimation      &   3.1\%     & ---    \\       
MC statistics		       &   2.2\%     &  3 -- 24\% \\
\end{tabular}
\end{ruledtabular}
\caption{Effect of the systematic uncertainties in the  two muon and
	 two jet sample on background and signal cross sections. The muon ID
	 contribution comprises the uncertainties due to muon reconstruction,
	 isolation, track finding and matching, and resolution for the two
	 muons. The systematic uncertainties on the signal strongly depend on the
	 neutralino mass, so a typical range is given.}
\label{table:syserrors}
\end{table}

%% Limits ------------------------------------------------------------------------------- 
In the absence of an excess in the data, we set cross section limits on
resonant slepton production. To be as model independent as possible, we
calculated 95\%~C.L. limits with respect to the slepton production
cross section times branching fraction to gaugino plus muon using the CL$_{s}$
method \cite{junk}. The limit is then given in the slepton-mass and
gaugino-mass plane, as shown in Fig.~\ref{fig:limxsecch1}.  
In addition, our results are shown in Fig.~\ref{fig:limldcomb} as
$\lambda'_{211}$ exclusion contours interpreted within the mSUGRA framework,
with $\tan\beta=5$, $\mu<0$, and $A_0=0$. The slepton-mass and
gaugino-mass pair define the universal scalar and fermion masses $m_0$ and
$m_{1/2}$. All three channels were combined to form one limit for
$q\bar{q}\to\tilde{l}$, with $\tilde{l}=\tilde{\mu},\tilde{\nu}_\mu$. 

A lower limit on the slepton mass for a given \lqd-coupling
$\lambda'_{211}$ can be extracted from Fig.~\ref{fig:limldcomb}. These limits
do not depend on other masses. They are indicated by arrows and summarized in
Table~\ref{tab:abs_lim_sl_sq}. Similarly, the exclusion contour can be
translated within mSUGRA into constraints on other masses and parameters.
\begin{table}[h]
  \begin{ruledtabular}
  \begin{tabular}{cclc}
        &\multicolumn{1}{c}{Excluded slepton mass range}        & \multicolumn{1}{c}{Coupling strength}& \\ \hline
        &  $m$($\tilde{l}$)    $\le\,210$~GeV & for $\lambda'_{211}\,\ge\,0.04$& \hfill \\
	&  $m$($\tilde{l}$)    $\le\,340$~GeV & for $\lambda'_{211}\,\ge\,0.06$& \\
	&  $m$($\tilde{l}$)    $\le\,363$~GeV & for $\lambda'_{211}\,\ge\,0.10$& \\
  \end{tabular}
  \end{ruledtabular}  
  \caption{ Limits on the slepton mass $\tilde{l}$ 
            for a given \lqd-coupling $\lambda'_{211}$ and $\tan\beta=5$,
	    $\mu<0$ from Fig.~\ref{fig:limldcomb}.}
  \label{tab:abs_lim_sl_sq}
\end{table}

In summary, we have searched for $R$-parity violating supersymmetry via a
non-zero \lqd-coupling $\lambda'_{211}$ in final states with at least two muons
and two jets. No excess in comparison with SM expectation was found and we set
model independent cross section limits, improved compared to D0 Run~I by one
order of magnitude. The limits are interpreted within the mSUGRA framework and
translated into the best constraints to date on the coupling strength
$\lambda'_{211}$.  D0 Run~I excluded slepton masses up to $280$~GeV for
$\lambda'_{211}=0.09$ and $m(\tilde{\chi}^0_1)=\nolinebreak200$~GeV. Now,
slepton masses up to $358$~GeV can be excluded, for $\lambda'_{211}=0.09$
independent of other masses.

%% Acknowledgement ------------------------------------------------------------------------
% acknowledgement_paragraph_r2.tex                4/27/06
%
We thank the staffs at Fermilab and collaborating institutions, 
and acknowledge support from the 
DOE and NSF (USA);
CEA and CNRS/IN2P3 (France);
FASI, Rosatom and RFBR (Russia);
CAPES, CNPq, FAPERJ, FAPESP and FUNDUNESP (Brazil);
DAE and DST (India);
Colciencias (Colombia);
CONACyT (Mexico);
KRF and KOSEF (Korea);
CONICET and UBACyT (Argentina);
FOM (The Netherlands);
PPARC (United Kingdom);
MSMT (Czech Republic);
CRC Program, CFI, NSERC and WestGrid Project (Canada);
BMBF and DFG (Germany);
SFI (Ireland);
The Swedish Research Council (Sweden);
Research Corporation;
Alexander von Humboldt Foundation;
and the Marie Curie Program.
%
   % input acknowledgement
%\\ 

%% Bibliography ---------------------------------------------------------------------------

\end{document}